\documentclass[aps,prc,preprint,showpacs,superscriptaddress,amsmath,amssymb]{revtex4}
\usepackage{graphicx}
\usepackage{dcolumn}
\usepackage{bm}
\usepackage{color}
\usepackage{booktabs, longtable}
\usepackage{multirow}
\usepackage{ltxtable}

\begin{document}
\title{Odd-even staggering of neutron radii for neutron-rich Mg isotopes in continuum Skyrme-Hartree-Fock-Bogoliubov theory}
\author{T. T. Sun}
\affiliation{School of Physics and State Key Laboratory of Nuclear Physics and Technology, Peking University, 100871 Beijing, People's Republic of China}
\affiliation{Graduate School of Science and Technology, Niigata University, Niigata 950-2181, Japan}
\author{M. Matsuo}
\affiliation{Graduate School of Science and Technology, Niigata University, Niigata 950-2181, Japan}
\affiliation{Department of Physics, Faculty of Science, Niigata University, Niigata 950-2181, Japan}
\author{Y. Zhang}
\affiliation{Graduate School of Science and Technology, Niigata University, Niigata 950-2181, Japan}
\affiliation{Department of Physics, Faculty of Science, Niigata University, Niigata 950-2181, Japan}
\author{J. Meng}
\email{mengj@pku.edu.cn}
 \affiliation{School of Physics and State Key Laboratory of Nuclear Physics and Technology, Peking University, 100871 Beijing, People's Republic of China}
 \affiliation{School of Physics and Nuclear Energy Engineering, Beihang University, 100191 Beijing, People's Republic of China}
 \affiliation{Department of Physics, University of Stellenbosch, 7602 Stellenbosch, South Africa}

\date{\today}

\begin{abstract}

The self-consistent continuum Skyrme-Hartree-Fock-Bogoliubov theory formulated with Green's function technique in the coordinate space is developed to investigate odd-$A$ nuclei by incorporating the blocking effect. In a calculation performed with the SLy4 parameter for the neutron-rich Mg isotopes with $A=36-42$ around the neutron shell closure $N=28$, the odd-even staggering in the neutron rms radius, i.e., a larger value in $^{39}$Mg than those in $^{38}$Mg and $^{40}$Mg, is found. The large neutron radius in $^{39}$Mg is due to the blocking effect on the pair correlation energy, for which the configuration occupying the weakly-bound quasiparticle state $2p_{3/2}$ becomes the ground state instead of the $1f_{7/2}$ configuration. Performing systematic calculations with different Skyrme parameters we find that the ground state configuration for the odd-$A$ Mg isotopes, and hence the odd-even staggering of neutron radii, are sensitive to the details in the single-particle spectra, especially the gap between orbits $1f_{7/2}$ and $2p_{3/2}$.
\end{abstract}

\maketitle

\section{Introduction}\label{Chapter1}
The nuclear halo phenomenon in neutron-rich nuclei near the drip-line has been considered as one of the most fascinating topics with many new and interesting features, such as very large rms matter radius as compared to that by $A^{1/3}$ law~\cite{PRL1985Tanihata55, ARNPS1995Hansen45, PPNP2000Casten45, PhyRep2004Jonson389, RevModPhys2004Jensen76}. The pairing correlation and the coupling with the continuum are found to be very important~\cite{PRC1996DobaczewskiJ53, PLB2000Bennaceur496, PRL1996Meng77, PPNP2006MengJ}, requiring that the theory for neutron-rich nuclei can properly deal with this two effects.

A useful tool for studying exotic nuclei is the Bogoliubov theory in the coordinate space with unified description of both the mean field and the pairing field. It has been applied to the Skyrme-Hartree-Fock-Bogoliubov (HFB) theory~\cite{NPA1984Doba422, PRC1996DobaczewskiJ53} and the relativistic Hartree-Bogoliubov (RHB) theory~\cite{PRL1996Meng77,PRL1998MengJ80,NPA1998MengJ, PPNP2006MengJ, PhysRep2005Vretenar409}. Besides, the deformed relativistic Hartree-Bogoliubov (DRHB) theory based on a spherical Woods-Saxon basis has also been developed to study the halo phenomenon in deformed nuclei~\cite{PRC2010Zhou82, PRC2012LLLI85, CPL2012Li29, PRC2012ChenY85}.

In many calculations, the box boundary condition is adopted for solving the H(F)B equations in the coordinate space, and hence the discretized quasiparticle states are obtained~\cite{PRC1996DobaczewskiJ53, PLB2000Bennaceur496, PRL1996Meng77, PPNP2006MengJ}. Although it is appropriate for deeply bound states, the box boundary condition is not suitable for weekly bound and continuum states unless a large enough box is taken. On the other hand, the Green's function method~\cite{YadFiz1987Belyaev45} has a merit to impose the correct asymptotic behaviors on the wave functions especially for the weakly bound and continuum states, and to calculate the densities.

The HFB theory with the Green's function method has been formulated for even-even nuclei~\cite{Shlomo1975507,NPA2001Matsuo696,PRC2009Oba80, PRC2011ZhangY83, PRC2012ZhangY86}. In 2011, Zhang et al. introduced the Green's function method to the self-consistent Skyrme-HFB theory~\cite{PRC2011ZhangY83}. In the present work, we extend the continuum Skyrme-HFB theory with Green's function method to discuss odd-$A$ nuclei by incorporating the blocking effect.

An interesting feature found in odd-$A$ isotopes is the odd-even staggering of the reaction cross section $\sigma_{R}$ and the interaction cross section $\sigma_{I}$~\cite{PRC2004FangDQ69, NPA2001Ozawa691, PLB2012TakechiNe}. Namely, the reaction (interaction) cross sections for neutron-rich odd-$N$ isotopes are enhanced as compared to the neighboring even-$N$ nuclei, indicating the staggering in the neutron radii~\cite{arXiv1305v1}. The phenomenon is found in $^{14-16}$C~\cite{PRC2004FangDQ69}, $^{18-20}$C~\cite{NPA2001Ozawa691}, and $^{28-32}$Ne~\cite{PLB2012TakechiNe}. In Ref.~\cite{PRC2011Hagino84}, the odd-even staggering in $^{30-32}$Ne and $^{14-16}$C are studied with the HFB method and a three-body model respectively, and it is attributed to the pairing anti-halo effect~\cite{PLB2000Bennaceur496}, by considering pairing effect in even-$N$ nuclei and vanishing it in odd-$N$ isotopes. However, the pairing correlations in odd-$N$ nuclei should be treated in the same self-consistent scheme as even-$N$ nuclei, together with the blocking effect~\cite{ManybodyProb2000} caused by the last neutron. Attempts to treat odd-A nucleus for Na isotopes~\cite{PLB1998JMeng419} and C, N, O and F isotopes~\cite{PLB2002Meng532} by the relativistic Hartree-Bogoliubov method and for C isotopes by the relativistic Hartree-Fock-Bogoliubov method~\cite{PRC2013LuXL87} have been reported.

In this paper, we will develop a self-consistent continuum Skyrme-HFB theory for odd-A nuclei formulated with Green's function technique in the coordinate space and explore the odd-even staggering of the neutron radii for neutron-rich Mg isotopes newly investigated experimentally~\cite{PComOhtsubo}. We focus on the mechanisms of odd-even staggering and the influences of the blocking effect on the pairing energy. In Sec.~\ref{Chapter2}, we introduce the formulation of the continuum Skyrme-HFB theory for odd-$A$ nuclei using the Green's function technique. Numerical details will be presented in Sec.~\ref{Chapter3}. After giving the results and discussions in Sec.~\ref{Chapter4}, finally conclusions are drawn in Sec.~\ref{Chapter5}.

\section{THEORETICAL FRAMEWORK}\label{Chapter2}

\subsection{Coordinate-space Hartree-Fock-Bogoliubov theory}

In the Hartree-Fock-Bogoliubov (HFB) theory, the pair correlated nuclear system is described in terms of independent quasiparticles introduced through the Bogoliubov transformation~\cite{ManybodyProb2000}. The quasiparticle states are solutions of the HFB equation which is written in the coordinate space representation~\cite{NPA1984Doba422} as
\begin{eqnarray}
\int d \bm{r'}\sum_{\sigma'}\left(
                               \begin{array}{cc}
                                 h(\bm{r}\sigma, \bm{r'}\sigma')-\lambda\delta({\bm r}-{\bm r'})\delta_{\sigma\sigma'}
                               & \tilde{h}({\bm r \sigma}, {\bm r'}\sigma')\\
                                 \tilde{h}^{*}({\bm r}\tilde{\sigma}, {\bm r'}\tilde{\sigma}')
                               & -h^{*}({\bm r}\tilde{\sigma}, {\bm r}'\tilde{\sigma}')+\lambda\delta({\bm r}-{\bm r}')\delta_{\sigma\sigma'} \\
                               \end{array}
                             \right)
                             \phi_{i}(\bm{r}'\sigma')&&\nonumber\\
                        =E_{i}\phi_{i}(\bm{r}\sigma),&&\label{EQ:HFBeq}
\end{eqnarray}
where $E_{i}$ is the quasiparticle energy, $\lambda$ is the Fermi energy and the notations follow Ref.~\cite{NPA2001Matsuo696}. The quasiparticle wave function $\phi_{i}(\bm{r}\sigma)$ and its conjugate wave function $\bar{\phi}_{\tilde{i}}(\bm{r}\sigma)$ have two components:
\begin{equation}
 \phi_{i}(\bm{r}\sigma)\equiv
   \left(
     \begin{array}{c}
       \varphi_{1,i}(\bm{r}\sigma) \\
       \varphi_{2,i}(\bm{r}\sigma) \\
     \end{array}
   \right),~~~~
 \bar{\phi}_{\tilde{i}}(\bm{r}\sigma)\equiv
   \left(
     \begin{array}{c}
       -\varphi_{2,i}^{*}(\bm{r}\tilde{\sigma}) \\
        \varphi_{1,i}^{*}(\bm{r}\tilde{\sigma}) \\
     \end{array}
  \right),
  \label{EQ:qpwf}
\end{equation}
where $\varphi(\bm{r}\tilde{\sigma})\equiv -2\sigma\varphi(\bm{r}, -\sigma)$. The Hartree-Fock Hamiltonian $h(\bm{r}\sigma, \bm{r'}\sigma')$ and the pair Hamiltonian $\tilde{h}(\bm{r}\sigma, \bm{r'}\sigma')$ can be obtained by the variation of the total energy functional with respect to the particle density matrix $\rho(\bm{r}\sigma, \bm{r}'\sigma')\equiv \langle\Phi_{0}|c^{\dag}_{\bm{r}'\sigma'}c_{\bm{r}\sigma}|\Phi_{0}\rangle$ and pair density matrix $\tilde{\rho}(\bm{r}\sigma, \bm{r}'\sigma')\equiv \langle\Phi_{0}|c_{\bm{r}'\tilde{\sigma}'}c_{\bm{r}\sigma}|\Phi_{0}\rangle$, respectively. Here $c_{\bm{r}\sigma}, c_{\bm{r}\sigma}^{\dag}$ are the particle operators and $|\Phi_{0}\rangle$ is the ground state of the system. The two density matrices can be combined in a generalized density matrix $\mathcal{R}$ as
\begin{equation}
 \mathcal{R}(\bm{r}\sigma,\bm{r'}\sigma')\equiv
  \left(
   \begin{array}{cc}
     \rho(\bm{r}\sigma, \bm{r}'\sigma')
    &\tilde{\rho}(\bm{r}\sigma, \bm{r}'\sigma') \\
     \tilde{\rho}^{*}(\bm{r}\tilde{\sigma}, \bm{r}'\tilde{\sigma}')
    &\delta({\bm r}-{\bm r'})\delta_{\sigma\sigma'}-\rho^{*}(\bm{r}\tilde{\sigma}, \bm{r}'\tilde{\sigma}') \\
   \end{array}
  \right),
  \label{EQ:Rmarix}
\end{equation}
where the particle density matrix $\rho(\bm{r}\sigma, \bm{r}'\sigma')$ and pair density matrix $ \tilde{\rho}(\bm{r}\sigma, \bm{r}'\sigma')$ are the $``11"$ and $``12"$ components of $\mathcal{R}(\bm{r}\sigma,\bm{r'}\sigma')$, respectively.

For an even-even nucleus, the ground state $|\Phi_{0}\rangle$ is represented as a vacuum with
respect to quasiparticles~\cite{ManybodyProb2000}, i.e.,
\begin{equation}
 \beta_{i}|\Phi_{0}\rangle=0,~~{\rm for~all}~~i=1,\cdots, M,
\end{equation}
where $\beta_{i}$ and $\beta_{i}^{\dag}$ are the quasiparticle annihilation and creation operators and $M$ is the dimension of the quasiparticle space. With the quasiparticle vacuum $|\Phi_{0}\rangle$, the generalized density matrix of Eq.~(\ref{EQ:Rmarix}) can also be expressed as
\begin{equation}
\mathcal{R}(\bm{r}\sigma,\bm{r}'\sigma')=\sum\limits_{i}\overline{\phi}_{\tilde{i}}(\bm{r}\sigma)\overline{\phi}^{\dag}_{\tilde{i}}(\bm{r}'\sigma').
\label{EQ:Reven}
\end{equation}

\subsection{Blocking effect for odd-$A$ nuclei}

For an odd-$A$ nucleus£¬the ground state is a one-quasiparticle state~$|\Phi_{1}\rangle$~\cite{ManybodyProb2000}, which can be constructed based on a HFB vacuum~$|\Phi_{0}\rangle$ as
\begin{equation}
   |\Phi_{1}\rangle=\beta_{i_{\rm b}}^{\dag}|\Phi_{0}\rangle,
   \label{EQ:WFodd}
\end{equation}
where $i_{\rm b}$ denotes the quantum number of the blocked quasiparticle state.

For the one-quasiparticle state $|\Phi_{1}\rangle$, the particle density matrix $\rho(\bm{r}\sigma,\bm{r}'\sigma')$ and pair density matrix $\tilde{\rho}(\bm{r}\sigma,\bm{r}'\sigma')$
\begin{subequations}
 \begin{eqnarray}
  \rho(\bm{r}\sigma,\bm{r}'\sigma')&\equiv\langle\Phi_{1}|c^{\dag}_{\bm{r}'\sigma'}c_{\bm{r}\sigma}|\Phi_{1}\rangle,\\
  \tilde{\rho}(\bm{r}\sigma,\bm{r}'\sigma')&\equiv\langle\Phi_{1}|c_{\bm{r}'\tilde{\sigma}'}c_{\bm{r}\sigma}|\Phi_{1}\rangle,
 \end{eqnarray}
\end{subequations}
and the generalized density matrix~$\mathcal{R}(\bm{r}\sigma,\bm{r}'\sigma')$ becomes
\begin{subequations}
 \begin{eqnarray}
  &&\mathcal{R}(\bm{r}\sigma,\bm{r}'\sigma')=\mathcal{R}_0(\bm{r}\sigma,\bm{r}'\sigma')-\mathcal{R}_1(\bm{r}\sigma,\bm{r}'\sigma')+\mathcal{R}_2(\bm{r}\sigma,\bm{r}'\sigma'),\label{EQ:ROdd}\\
  &&\mathcal{R}_0(\bm{r}\sigma,\bm{r}'\sigma')=\sum\limits_{i:{\rm all}}\overline{\phi}_{\tilde{i}}(\bm{r}\sigma)\overline{\phi}^{\dag}_{\tilde{i}}(\bm{r}'\sigma'),
  \label{EQ:ROdd-1}\\
  &&\mathcal{R}_1(\bm{r}\sigma,\bm{r}'\sigma')=\overline{\phi}_{\tilde{i}_{\rm b}}(\bm{r}\sigma)\overline{\phi}^{\dag}_{\tilde{i}_{\rm b}}(\bm{r}'\sigma'),
  \label{EQ:ROdd-2}\\
  && \mathcal{R}_2(\bm{r}\sigma,\bm{r}'\sigma')=\phi_{i_{\rm b}}(\bm{r}\sigma)\phi_{i_{\rm b}}^{\dag}(\bm{r}'\sigma').
  \label{EQ:ROdd-3}
 \end{eqnarray}
 \label{EQ:Rodd}
\end{subequations}
Compared with the generalized density matrix $\mathcal{R}(\bm{r}\sigma,\bm{r}'\sigma')$ of Eq.~(\ref{EQ:Reven}) for even-even nuclei, two more terms $\mathcal{R}_1(\bm{r}\sigma,\bm{r}'\sigma')$ and $\mathcal{R}_2(\bm{r}\sigma,\bm{r}'\sigma')$ are introduced for the odd$-A$ nuclei because of the blocking effect.

\subsection{Densities for odd-$A$ nuclei using the Green's function method}

The particle density $\rho({\bm r}\sigma, {\bm r}'\sigma')$ and pair density $\tilde{\rho}({\bm r}\sigma, {\bm r}'\sigma')$ for odd$-A$ nuclei are given as
\begin{subequations}
 \begin{eqnarray}
 \rho({\bm r}\sigma, {\bm r}'\sigma')&=& \rho_{0}({\bm r}\sigma, {\bm r}'\sigma')- \rho_{1}({\bm r}\sigma, {\bm r}'\sigma')+ \rho_{2}({\bm r}\sigma, {\bm r}'\sigma')\nonumber\\
 &=&\sum_{i:{\rm all}}\varphi_{2,i}^{*}({\bm r}\tilde{\sigma})\varphi_{2,i}({\bm r}'\tilde{\sigma}')-\varphi_{2,i_{\rm b}}^{*}({\bm r}\tilde{\sigma})\varphi_{2,i_{\rm b}}({\bm r}'\tilde{\sigma}')+\varphi_{1,i_{\rm b}}({\bm r}\sigma)\varphi^{*}_{1,i_{\rm b}}({\bm r}'\sigma'), \label{EQ:rho-odd} \\
 \tilde{\rho}({\bm r}\sigma, {\bm r}'\sigma')&=& \tilde{\rho}_{0}({\bm r}\sigma, {\bm r}'\sigma')-\tilde{ \rho}_{1}({\bm r}\sigma, {\bm r}'\sigma')+\tilde{ \rho}_{2}({\bm r}\sigma, {\bm r}'\sigma')\nonumber\\
 &=&\sum_{i:{\rm all}}\varphi_{2,i}^{*}({\bm r}\tilde{\sigma})\varphi_{1,i}({\bm r}'\tilde{\sigma}')-\varphi_{2,i_{\rm b}}^{*}({\bm r}\tilde{\sigma})\varphi_{1,i_{\rm b}}({\bm r}'\tilde{\sigma}')-\varphi_{1,i_{\rm b}}({\bm r}\sigma)\varphi^{*}_{2,i_{\rm b}}({\bm r}'\sigma').
 \label{EQ:rhot-odd}
 \end{eqnarray}
 \label{EQ:rhorhot}
\end{subequations}

\begin{figure}[!ht]
  \includegraphics[width=0.45\textwidth]{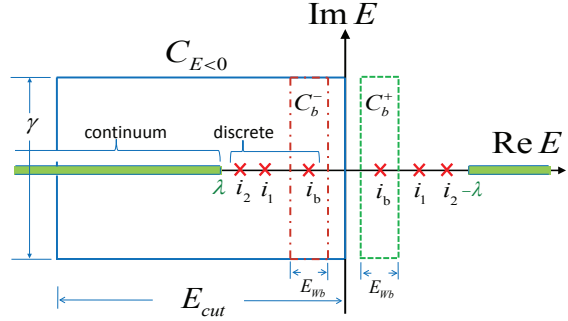}\\
  \caption{(Color online) Contour paths $C_{E<0}, C_{\rm b}^{-}, C_{\rm b}^{+}$ to perform the integrations of the Green's function on the complex quasiparticle energy plane. The paths are chosen to be rectangles with the same width $\gamma$ and different lengths, i.e., $E_{\rm cut}$, $E_{\rm Wb}$, and $E_{\rm Wb}$ for $C_{E<0}, C_{\rm b}^{-}$, and $C_{\rm b}^{+}$ respectively. The crosses denote the discrete quasiparticle states and the continuum states are denoted by the thick solid line.}
  \label{Fig1}
\end{figure}

We now rewrite Eqs.~(\ref{EQ:Rodd}) and (\ref{EQ:rhorhot}) using the HFB Green's function. The spectral representation of the HFB Green's function is expressed as~\cite{YadFiz1987Belyaev45}
\begin{equation}
  G({\bm r}\sigma, {\bm r}'\sigma', E)
 =\sum_{i}\left(\frac{\phi_{i}({\bm r}\sigma)\phi_{i}^{\dag}({\bm r}'\sigma')}{E-E_{i}}+\frac{\bar{\phi}_{\tilde{i}}({\bm r}\sigma)\bar{\phi}_{\tilde{i}}^{\dag}({\bm r}'\sigma')}{E+E_{i}}\right),
 \label{EQ:GF}
\end{equation}
which has two branches. One branch is related with the single quasiparticle wave function $\phi_{i}({\bm r}\sigma)$ and positive eigenvalues $E_{i}$, and the other one is related with the single quasiparticle conjugate wave function $\bar{\phi}_{\tilde{i}}({\bm r}\sigma)$ and negative eigenvalues $-E_{i}$. According to the Cauchy's theorem, the terms $\mathcal{R}_{0}({\bm r}\sigma,{\bm r}'\sigma')$, $\mathcal{R}_{1}({\bm r}\sigma,{\bm r}'\sigma')$, and $\mathcal{R}_{2}({\bm r}\sigma,{\bm r}'\sigma')$ in Eq.~(\ref{EQ:Rodd}) for the generalized density matrix can be calculated with the integrals of the Green's function in the complex quasiparticle energy plane as
\begin{subequations}
 \begin{eqnarray}
  &&\mathcal{R}_{0}({\bm r}\sigma,{\bm r}'\sigma')=\frac{1}{2\pi i}\oint_{C_{E<0}} dE G({\bm r}\sigma,{\bm r}'\sigma',E),\\
  &&\mathcal{R}_{1}({\bm r}\sigma,{\bm r}'\sigma')=\frac{1}{2\pi i}\oint_{C_{\rm b}^{-}} dE G({\bm r}\sigma,{\bm r}'\sigma',E),\\
  &&\mathcal{R}_{2}({\bm r}\sigma,{\bm r}'\sigma')=\frac{1}{2\pi i}\oint_{C_{\rm b}^{+}} dE G({\bm r}\sigma,{\bm r}'\sigma',E),
  \label{R012}
 \end{eqnarray}
\end{subequations}
where the contour path $C_{E<0}$ encloses all the negative quasiparticle energies $-E_{i}$, $C_{\rm b}^{-}$ encloses only the pole $-E_{i_{\rm b}}$ and $C_{\rm b}^{+}$ encloses only the pole $E_{i_{\rm b}}$, which can be seen in Fig.~\ref{Fig1}.

From now on, we assume that the system is spherical and apply a filling approximation, i.e., we take an average of the blocked quasiparticle state $i_{\rm b}=n_{\rm b}l_{\rm b}j_{\rm b}m_{j_{\rm b}}$ over the magnetic quantum numbers $m_{j_{\rm b}}= -j_{\rm b}, -j_{\rm b}+1, \cdots, j_{\rm b}-1, j_{\rm b}$.
The quasiparticle wave function $\phi_{i}({\bm r}\sigma)$, the generalized density matrix $\mathcal{R}({\bm r}\sigma,{\bm r}'\sigma')$, and the Green's function $G({\bm r}\sigma,{\bm r}'\sigma')$ can be expanded using the spinor spherical harmonics as
\begin{subequations}
\begin{eqnarray}
\phi_{i}({\bm r}\sigma)&=&\frac{1}{r}\phi_{lj}(r)Y_{ljm}(\hat{\bm r}\sigma),~~~ {\rm where}~~~
 \phi_{lj}(r)=\left(
                  \begin{array}{c}
                    \varphi_{1,lj}(r) \\
                    \varphi_{2,lj}(r) \\
                  \end{array}
             \right),\\
 \mathcal{R}({\bm r}\sigma, {\bm r'}\sigma')&=&\sum_{ljm}Y_{ljm}(\hat{\bm r}\sigma)\mathcal{R}_{lj}(r,r')Y_{ljm}^{*}(\hat{\bm r}'\sigma'),\\
  G({\bm r}\sigma, {\bm r'}\sigma')&=&\sum_{ljm}Y_{ljm}(\hat{\bm r}\sigma)\frac{\mathcal{G}_{0,lj}(r,r')}{rr'}Y_{ljm}^{*}(\hat{\bm r}'\sigma').
\end{eqnarray}
\end{subequations}
As a result, the radial part of the local particle density $\rho({\bm r})=\sum\limits_{\sigma}\rho({\bm r}\sigma,{\bm r}\sigma)=\sum\limits_{\sigma}\mathcal{R}^{11}({\bm r}\sigma,{\bm r}\sigma)$ and pair density $\tilde{\rho}({\bm r})=\sum\limits_{\sigma}\tilde{\rho}({\bm r}\sigma, {\bm r}\sigma)=\sum\limits_{\sigma}\mathcal{R}^{12}({\bm r}\sigma, {\bm r}\sigma)$ can be expressed by the radial box-discretized quasiparticle wave functions $\phi_{nlj}(r)$ or the radial HFB Green's function $\mathcal{G}_{0,lj}(r,r')$ as
\begin{subequations}
\begin{eqnarray}
\rho(r)&=&\frac{1}{4\pi}\sum_{lj:{\rm all}}(2j+1)\mathcal{R}_{0,lj}^{11}(r,r)-\frac{1}{4\pi}\mathcal{R}_{1,l_{\rm b}j_{\rm b}}^{11}(r,r)+\frac{1}{4\pi}\mathcal{R}_{2,l_{\rm b}j_{\rm b}}^{11}(r,r)\nonumber\\
&=&\frac{1}{4\pi r^{2}}\left[\sum_{lj:{\rm all}}(2j+1)\sum_{n:{\rm all}}\varphi_{2,nlj}^{2}(r)-\varphi_{2,n_{\rm b}l_{\rm b}j_{\rm b}}^{2}(r)+\varphi_{1,n_{\rm b}l_{\rm b}j_{\rm b}}^{2}(r)\right]\nonumber\\
&=&\frac{1}{4\pi r^{2}}\frac{1}{2\pi i}\left[\sum_{lj:{\rm all}}(2j+1)\oint_{C_{E<0}}dE\mathcal{G}_{0,lj}^{11}(r,r,E)\right.\nonumber\\
&&\left.~~~~~~~~~~~~~~~~~~~~~-\oint_{C^{-}_{\rm b}}dE\mathcal{G}_{0,l_{\rm b}j_{\rm b}}^{11}(r,r,E)+\oint_{C^{+}_{\rm b}}dE\mathcal{G}_{0,l_{\rm b}j_{\rm b}}^{11}(r,r,E)\right],\label{EQ:rho}\\
\tilde{\rho}(r)&=&\frac{1}{4\pi}\sum_{lj:{\rm all}}(2j+1)\mathcal{R}_{0,lj}^{12}(r,r)-\frac{1}{4\pi}\mathcal{R}_{1,l_{\rm b}j_{\rm b}}^{12}(r,r)+\frac{1}{4\pi}\mathcal{R}_{2,l_{\rm b}j_{\rm b}}^{12}(r,r)\nonumber\\
&=&\frac{1}{4\pi r^{2}}\left[\sum_{lj:{\rm all}}(2j+1)\sum_{n:{\rm all}}\varphi_{1,nlj}(r)\varphi_{2,nlj}(r)\right.\nonumber\\
&&\left.~~~~~~~~~~~~~~~~~~~~~-\varphi_{1,n_{\rm b}l_{\rm b}j_{\rm b}}(r)\varphi_{2,n_{\rm b}l_{\rm b}j_{\rm b}}(r)-\varphi_{2,n_{\rm b}l_{\rm b}j_{\rm b}}(r)\varphi_{1,n_{\rm b}l_{\rm b}j_{\rm b}}(r)\right]\nonumber\\
&=&\frac{1}{4\pi r^{2}}\frac{1}{2\pi i}\left[\sum_{lj:{\rm all}}(2j+1)\oint_{C_{E<0}}dE\mathcal{G}_{0,lj}^{12}(r,r,E)\right.\nonumber\\
&&\left.~~~~~~~~~~~~~~~~~~~~~-\oint_{C^{-}_{\rm b}}dE\mathcal{G}_{0,l_{\rm b}j_{\rm b}}^{12}(r,r,E)+\oint_{C^{+}_{\rm b}}dE\mathcal{G}_{0,l_{\rm b}j_{\rm b}}^{12}(r,r,E)\right].\label{EQ:rhot}
\end{eqnarray}
\end{subequations}
Here, we call the sum of the two terms $\varphi_{1,n_{\rm b}l_{\rm b}j_{\rm b}}(r)\varphi_{2,n_{\rm b}l_{\rm b}j_{\rm b}}(r)$ and $\varphi_{2,n_{\rm b}l_{\rm b}j_{\rm b}}(r)\varphi_{1,n_{\rm b}l_{\rm b}j_{\rm b}}(r)$ in Eq.~(\ref{EQ:rhot}) the blocking term for the pair density $\tilde{\rho}_{\rm b}(r)=2\varphi_{1,n_{\rm b}l_{\rm b}j_{\rm b}}(r)\varphi_{2,n_{\rm b}l_{\rm b}j_{\rm b}}(r)$. Similarly, one can express other radial local densities needed in the functional of the Skyrme interaction~\cite{NPA1975Engel249, RevModPhys2003Bender75}, such as the kinetic-energy density $\tau(r)$, the spin-orbit density $J(r)$, and etc., in terms of the radial Green's function.

To impose the correct boundary condition on the quasiparticle states, we replace the Green's function in the spectral representation, Eq.~(\ref{EQ:GF}), with the exact Green's function, in which the weakly bound and continuum states are treated exactly. In fact, the exact radial Green's function $\mathcal{G}_{0,lj}(r,r',E)$ can be constructed with the independent regular and outgoing solutions of the radial HFB equation with proper boundary conditions for the wave functions. For the outgoing solution, the quasiparticle wave function is connected at the box size $r=R$ to the asymptotic wave $\phi_{lj}^{(\rm out)}(r, E)/r=(Ah_{l}^{(+)}(k_{+}r), Bh_{l}^{(+)}(k_{-}r))^{T}$ with the spherical Hankel function $h_{l}^{(+)}(k_{\pm}r)$ and $k_{\pm}(E)=\sqrt{2m(\lambda\pm E)/\hbar^{2}}$. For other details, we refer the readers to Refs.~\cite{NPA2001Matsuo696, PRC2011ZhangY83}.

In the present work, we neglect possibility of deformation in Mg isotopes to focus on the pairing effect on the odd-even staggering in neutron rms radius.

\section{NUMERICAL DETAILS}\label{Chapter3}

In the $ph$ channel, we mainly use the Skyrme parameter SLy4~\cite{NPA1998ChabanatM635Skyrme}, but other parameter sets are also used for comparison. For the pairing interaction in the $pp$ channel, a density dependent $\delta$ interaction (DDDI) is adopted
\begin{equation}
 v_{\rm pair}({\bm r},{\bm r'})=\frac{1}{2}(1-P_{\sigma})V_0\left[1-\eta\left(\frac{\rho({\bm r})}{\rho_0}\right)^{\alpha}\right]\delta({\bm r}-{\bm r'}),
 \label{EQ:DDDI}
\end{equation}
with which the pair Hamiltonian $\tilde{h}({\bm r}\sigma, {\bm r}'\sigma')$ is reduced to the local pair potential~\cite{NPA1984Doba422}
\begin{equation}
 \Delta({\bm r})=\frac{1}{2}V_0\left[1-\eta\left(\frac{\rho({\bm r})}{\rho_0}\right)\right]\tilde{\rho}({\bm r}).
\end{equation}
The DDDI parameters in Eq.~(\ref{EQ:DDDI}) are taken as $V_{0}=-458.4$~MeV$\cdot$fm$^{3}$, $\eta=0.71$, $\alpha=0.59$, and $\rho_{0}=0.08$~fm$^{-3}$, with which the experimental neutron pairing gaps for the Sn isotopes are approximately reproduced~\cite{PRC2006MatsuoM73, NPA2007Matsuo788, PRC2010Matsuo82}. Furthermore, with the present pairing interaction strength $V_{0}$, the DDDI reproduces in the low density limit the scattering length $a=-18.5$~fm in the $^{1}S$ channel of the bare nuclear force~\cite{PRC2006MatsuoM73}. The cut-off of the quasiparticle states are taken with maximal angular momentum $j_{\rm max}=\frac{25}{2}$ and the maximal quasiparticle energy $E_{\rm cut}=60$~MeV.

To perform the integrals of the Green's function, the contour paths $C_{E<0}, C^{-}_{\rm b}, C^{+}_{\rm b}$ are chosen to be three rectangles on the complex quasiparticle energy plane as shown in Fig.~\ref{Fig1}, with the same width $\gamma=0.1$~MeV and different lengths, i.e., $E_{\rm cut}$, $E_{\rm Wb}$, $E_{\rm Wb}$ respectively. To enclose all the negative quasiparticle energies, the length of the contour path $C_{E<0}$ is taken as the maximal quasiparticle energy $E_{\rm cut}=60$~MeV. In the present discussions for Mg isotopes, the contour paths $C^{+}_{\rm b}$ and $C^{-}_{\rm b}$ are symmetric with respect to the origin and have the same length $E_{\rm Wb}=0.2~$MeV, which enclose the discrete quasiparticle states at $E_{i_{\rm b}}$ and $-E_{i_{\rm b}}$ in the center respectively. For the contour integration, we adopt an energy step $\Delta E=0.01$~MeV on the contour path. The HFB equation is solved with the box size $R=20$~fm and mesh size $\Delta r=0.1$~fm in the coordinate space.

The HFB iteration is performed until the convergence is achieved. In the iteration, we impose the particle number condition $\langle\Phi_{0}|\hat{N} |\Phi_{0}\rangle=N$ for even-even nuclei and $\langle\Phi_{1}|\hat{N} |\Phi_{1}\rangle=N$ for odd-$A$ nuclei.

\section{RESULTS AND DISCUSSION}\label{Chapter4}

In this section, neutron-rich Mg isotopes will be investigated by both the blocked continuum and box-discretized Skyrme-HFB approaches. We will focus our attentions on the odd-even staggering in the neutron radius and analyze its mechanism.

\subsection{Odd-even staggering in neutron rms radius}
\label{ResultA}

\begin{table}[!ht]
\tabcolsep=5pt
\caption{Blocked quasiparticle states $(njl)_{\rm b}$, total energy $E_{\rm tot.}$, two-neutron separation energy $S_{2{\rm n}}$, Fermi energy $\lambda$, neutron pairing energy $E_{\rm pair}^{\rm n}$, and neutron average pairing gap $\Delta$ for the Mg isotopes with $A=36-42$. Listed are also the Hartree-Fock (HF) single-particle energies $\varepsilon$ and single quasiparticle energies $E$ of $2p_{3/2}$ and $1f_{7/2}$ states for each isotope. The adopted Skyrme parameter is SLy4. Unit for energy is MeV.}
\label{Tab1}
\begin{tabular}{ccccccccc}
\hline\hline
    &$A$                   &   36   &    37    &   38   &    39    &    40  &   41     &   42   \\\hline
    &$(njl)_{\rm b}$        &   $-$  &$1f_{7/2}$&   $-$  &$2p_{3/2}$&    $-$ &$2p_{3/2}$&   $-$  \\
    &$E_{\rm tot.}$        &-261.563&-261.101  &-263.618&-263.169  &-265.099&-264.814  &-265.906\\
    &$S_{2{\rm n}}$        &   2.807&   2.457  &   2.036&   2.035  &   1.493&   1.726  &   0.801\\
    &$\lambda     $        &  -1.883&  -1.615  &  -1.500&  -1.534  &  -1.105&  -1.088  &  -0.697\\
    &$E_{\rm pair}^{\rm n}$& -10.420&  -6.169  &  -9.692&  -7.452  &  -7.086&  -2.629  &  -5.434\\
    &$\Delta$              &   1.553&   1.146  &   1.406&   1.268  &   1.103&   0.683  &   0.874\\\hline
\multirow{2}{*}{$2p_{3/2}$}
    &$\varepsilon$         &  -0.047&  -0.142  &  -0.254&  -0.354  &  -0.488&  -0.600  &  -0.771\\
    &$E $                  &   1.539&   1.632  &   1.554&   1.453  &   1.085&   0.743  &   0.805\\\hline
\multirow{2}{*}{$1f_{7/2}$}
    &$\varepsilon$         &  -1.697&  -1.910  &  -2.101&  -2.213  &  -2.516&  -2.690  &  -2.828\\
    &$E $                  &   1.855&   1.404  &   1.801&   1.646  &   1.949&   1.790  &   2.390\\
\hline\hline
\end{tabular}
\end{table}

Table~\ref{Tab1} lists the blocked quasiparticle states $(njl)_{\rm b}$, the total energy $E_{\rm tot.}$, the two-neutron separation energy $S_{2{\rm n}}(N,Z)=E_{\rm tot.}(N-2, Z)-E_{\rm tot.}(N, Z)$, the Fermi energy $\lambda$, the neutron pairing energy $E_{\rm pair}^{\rm n}$, and the neutron average pairing gap $\Delta$ for the Mg isotopes with mass number $A=36-42$ obtained by the continuum Skyrme-HFB calculations with SLy4 parameter. The odd$-N$ isotopes are calculated with blocking different neutron quasiparticle states around the Fermi surface and we choose the configuration with the lowest total energy. For $^{37}$Mg, $^{39}$Mg, and $^{41}$Mg, the blocked states $(nlj)_{\rm b}$ are found to be $1f_{7/2}$, $2p_{3/2}$, and $2p_{3/2}$ respectively. The neutron pairing energy and the neutron average pairing gap are calculated by
\begin{subequations}
\begin{eqnarray}
  E_{\rm pair}^{\rm n}&=&\frac{1}{2}\int d{\bm r}\Delta({\bm r})\tilde{\rho}({\bm r}),\\
  \Delta &=&\frac{\int d{\bm r}\Delta({\bm r})\tilde{\rho}({\bm r})}{\int d{\bm r}\tilde{\rho}({\bm r})},
\end{eqnarray}
\end{subequations}
weighted by the neutron pair density $\tilde{\rho}({\bm r})$. The nuclei listed in Table~\ref{Tab1} are bound as their two-neutron separation energies $S_{2{\rm n}}$ are positive. Note also that the blocked states in the odd-$N$ isotopes are bound, as their quasiparticle energies satisfy $E<-\lambda$. Except for nucleus $^{39}$Mg, the neutron pairing energy $E_{\rm pair}^{\rm n}$ and the average pairing gap $\Delta$ for odd$-A$ nuclei are smaller than those of the neighboring even-even nuclei. It is because the blocking term $\tilde{\rho}_{\rm b}=2\varphi_{1,n_{\rm b}l_{\rm b}j_{\rm b}}(r)\varphi_{2,n_{\rm b}l_{\rm b}j_{\rm b}}(r)$ in Eq.~(\ref{EQ:rhot}) reduces the pair correlation for odd$-N$ nuclei. Furthermore, because $N=28$ is a magic number, the neutron pair correlation in $^{40}$Mg is weaker than that of the neighboring nuclei.

\begin{figure}[!ht]
 \includegraphics[width=0.45\textwidth]{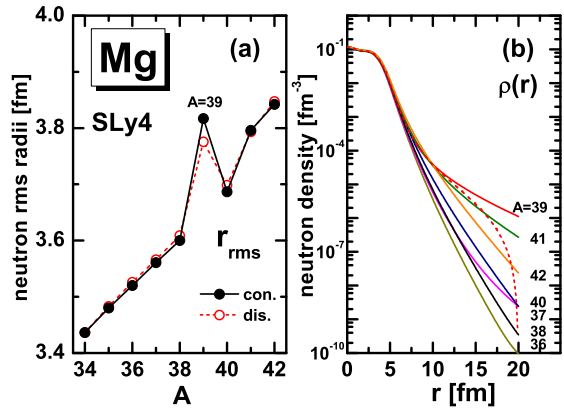}
 \caption{(Color online) (a) Neutron rms radius $r_{\rm rms}$ and (b) neutron density $\rho(r)$ for Mg isotopes. The filled circles connected with solid curves are the results of the continuum Skyrme-HFB calculations while the open circles with the dashed curve are those obtained in the box-discretized Skyrme-HFB calculations. The Skyrme parameter is SLy4.}
 \label{Fig2}
\end{figure}

Fig.~\ref{Fig2}(a) shows the neutron rms radius $r_{\rm rms}=[\int 4\pi r^{4}\rho(r)dr/\int 4\pi r^{2}\rho(r)dr]^{1/2}$ of Mg isotopes obtained in the continuum (filled circle) and box-discretized (open circle) Skyrme-HFB calculations with SLy4 parameter. It is found that $^{39}$Mg has a much larger rms radius than $^{38}$Mg and $^{40}$Mg, leading to a strong odd-even staggering. In Fig.~\ref{Fig2}(b) plotting the neutron density, we can see that the tail of the density distribution for $^{39}$Mg is very large and extended, and contributes to the large rms radius. Furthermore, we note also that the calculation with the box boundary condition underestimates the rms radius in $^{39}$Mg because of the inappropriate boundary condition, as shown in Fig.~\ref{Fig2}(b).

\begin{figure}[!ht]
 \includegraphics[width=0.45\textwidth]{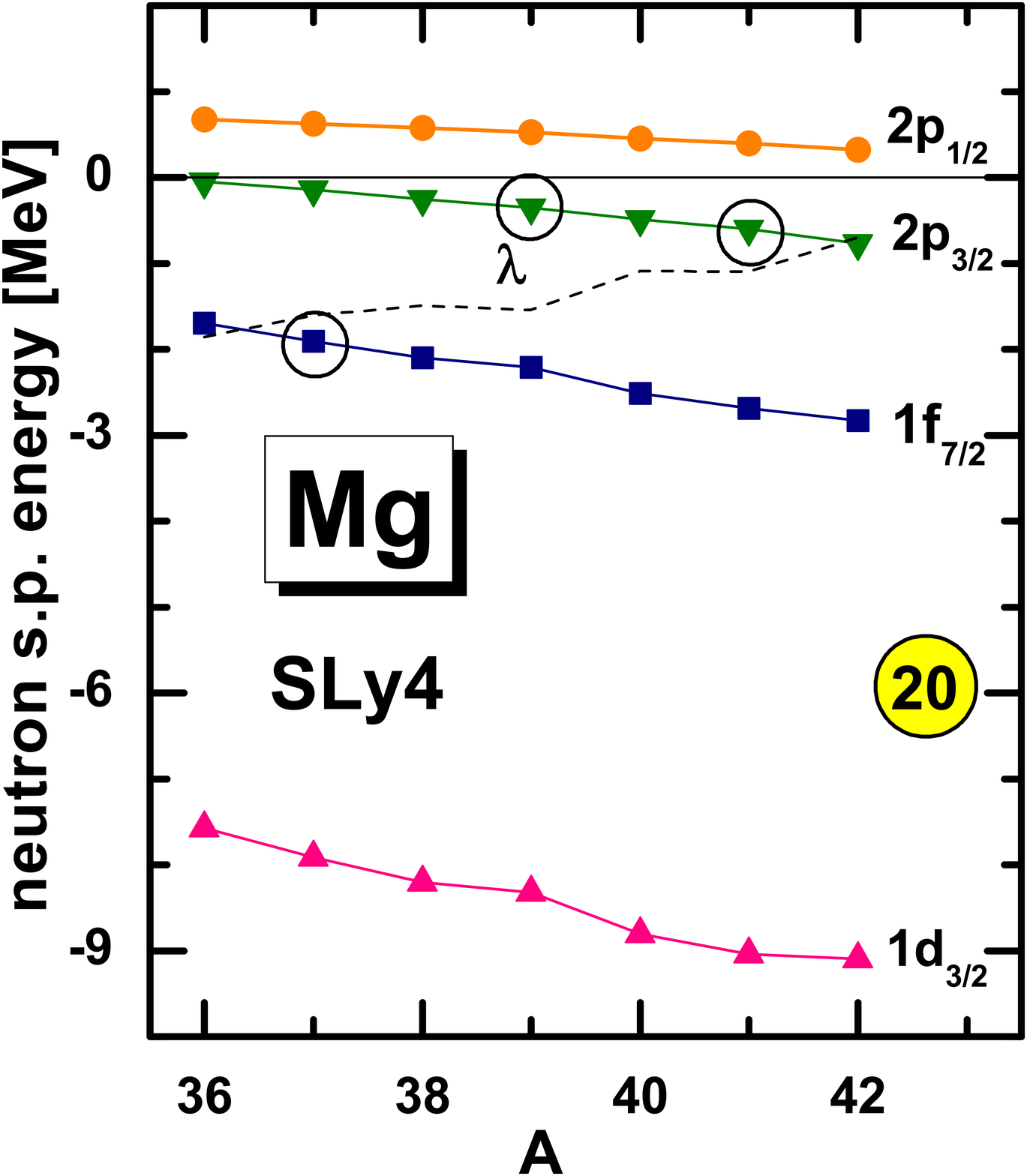}
 \caption{(Color online) Neutron Hartree-Fock single-particle energy $\varepsilon$ of Mg isotopes around the Fermi energy.  The dashed line denotes the neutron Fermi energy $\lambda$. The circled orbits correspond to the blocked quasiparticle states for odd$-A$ nuclei. The Skyrme parameter is SLy4.}
 \label{Fig3}
\end{figure}

We show in Fig.~\ref{Fig3} the neutron Fermi energy $\lambda$ as well as the Hartree-Fock (HF) single-particle energies $\varepsilon$, which are eigenenergies of the HF Hamiltonian $h$ (obtained after the final convergence of the blocked continuum Skyrme-HFB calculations). The single-particle orbits corresponding to the blocked quasiparticle states in odd$-A$ nuclei are labeled by circles. The values of the HF single-particle energies $\varepsilon$ are given in Table~\ref{Tab1}.

One can see from Fig.~\ref{Fig3} that as the mass number $A$ of Mg isotopes increases, the Fermi energy is raised up, to a position near the single-particle continuum threshold, while all the HF single-particle orbits fall down. Explicitly, $2p_{3/2}$ is the most weakly bound orbit and $1f_{7/2}$ is the second one. From $^{37}$Mg to $^{41}$Mg, the Fermi surface lies between $2p_{3/2}$ and $1f_{7/2}$ orbits.

\begin{figure}[!ht]
 \includegraphics[width=0.45\textwidth]{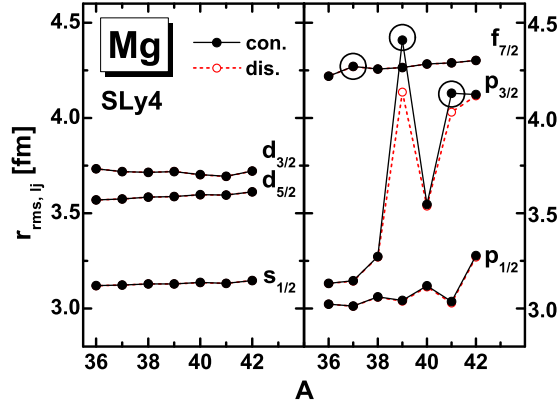}
 \caption{(Color online) Neutron rms radius $r_{{\rm rms},lj}$ of the $s, p, d$ and $f$ partial waves of Mg isotopes calculated for the $lj-$decomposed neutron density $\rho_{lj}(r)$. The filled symbols are the results obtained in the continuum Skyrme-HFB calculation with SLy4 parameter while the open symbols are the results obtained in the box-discretized Skyrme-HFB calculation. The circles indicate the blocked quasiparticle states.}
 \label{Fig4}
\end{figure}

To know the contributions to the total neutron rms radius from each state, we also calculate the rms radius for different partial waves $lj$,
\begin{equation}
 r_{{\rm rms},lj}=\left(\frac{\int 4\pi r^{4}\rho_{lj}(r)dr}{\int 4\pi r^{2}\rho_{lj}(r)dr}\right)^{1/2},
\end{equation}
weighted with the corresponding $lj$-decomposed neutron density
\begin{eqnarray}
\rho_{lj}(r)
  &=&\frac{1}{4\pi r^{2}}\left[(2j+1)\sum_{n:{\rm all}}\varphi_{2,nlj}^{2}(r)-\delta_{nlj,n_{\rm b}l_{\rm b}j_{\rm b}}\varphi_{2,n_{\rm b}l_{\rm b}j_{\rm b}}^{2}(r)+\delta_{nlj,n_{\rm b}l_{\rm b}j_{\rm b}}\varphi_{1,n_{\rm b}l_{\rm b}j_{\rm b}}^{2}(r)\right]\nonumber\\
  &=&\frac{1}{4\pi r^{2}}\frac{1}{2\pi i}\left[(2j+1)\oint_{C_{E<0}}dE\mathcal{G}_{0,lj}^{12}(r,r,E)\right.\nonumber\\
  & &\left.~~-\delta_{lj,l_{\rm b}j_{\rm b}}\oint_{C^{-}_{\rm b}}dE\mathcal{G}_{0,l_{\rm b}j_{\rm b}}^{12}(r,r,E)+\delta_{lj,l_{\rm b}j_{\rm b}}\oint_{C^{+}_{\rm b}}dE\mathcal{G}_{0,l_{\rm b}j_{\rm b}}^{12}(r,r,E)\right].
  \label{EQ:rholj}
\end{eqnarray}
In Fig.~\ref{Fig4}, we plot the rms radii for $lj=s_{1/2}, p_{1/2}, p_{3/2}, d_{3/2}, d_{5/2}$, and $f_{7/2}$ partial waves, obtained with the continuum Skyrme-HFB calculation (filled symbols) and the box-discretized Skyrme-HFB calculation (open symbols). The circles denote the blocked quasiparticle states. We can see clearly that the large rms radius of $^{39}$Mg compared with that of neighboring nuclei is mainly due to the large contribution from the $p_{3/2}$ partial wave. Meanwhile, the rms radii for other partial waves of Mg isotopes do not exhibit large odd-even staggering. Smaller neutron rms radius for the partial wave $p_{3/2}$ is obtained compared with $f_{7/2}$, because it includes also the deeply bound $1p_{3/2}$ state which reduces the total rms radius of the partial wave $p_{3/2}$.

In the box-discretized calculations, the particle density for each partial wave $lj$ is $\rho_{lj}(r)=\frac{2j+1}{4\pi r^{2}}\sum_{n}\varphi_{2,nlj}^{2}(r)$ if the partial wave $lj$ does not include the blocked state $l_{\rm b}j_{\rm b}$. For the blocked state, the upper and lower components of the single quasiparticle wave function $\varphi_{1,n_{\rm b}l_{\rm b}j_{\rm b}}(r)$ and $\varphi_{2,n_{\rm b}l_{\rm b}j_{\rm b}}(r)$ exchange. As a result, the density for the blocked orbit is $\varphi^{2}_{1,n_{\rm b}l_{\rm b}j_{\rm b}}(r)$  instead of $\varphi^{2}_{2,n_{\rm b}l_{\rm b}j_{\rm b}}(r)$. For the discrete quasiparticle states with quasiparticle energies $E<-\lambda$, the quasiparticle wave functions $\varphi_{1}(r)$ and $\varphi_{2}(r)$ in Eq.~(\ref{EQ:qpwf}) have different asymptotic behaviors: $\varphi_{1}(r)\rightarrow e^{-k_{+} r}$ with $k_{+}=\sqrt{\frac{2m|\lambda+E|}{\hbar^{2}}}$ and $\varphi_{2}(r)\rightarrow e^{-k_{-}r}$ with $k_{-}=\sqrt{\frac{2m|\lambda-E|}{\hbar^{2}}}$ for $r\rightarrow\infty$. Obviously, $\varphi_{1}(r)$ has a long tail with small $|\lambda+E|$. In Table~\ref{Tab1}, we list the quasiparticle energies $E$ of states $2p_{3/2}$ and $1f_{7/2}$ for each Mg isotope and it can be seen that the blocked states $1f_{7/2}$ for $^{37}$Mg, $2p_{3/2}$ for $^{39}$Mg, and $2p_{3/2}$ for $^{41}$Mg are bound. Explicitly, the one-quasiparticle state $2p_{3/2}$ for $^{39}$Mg is very weakly bound, where the quasiparticle energy $E=1.453$~MeV is very close to the threshold energy $-\lambda=1.534$~MeV for the continuum, i.e., $|\lambda+E|=0.081~$MeV. As a result, the density for $2p_{3/2}$ of $^{39}$Mg is very extended, leading to the large rms radius. Concerning $^{41}$Mg, the ground state is the configuration occupying the $2p_{3/2}$ state, but the odd-even staggering is weaker than that in $^{39}$Mg. The reason is that the $2p_{3/2}$ state of $^{41}$Mg is more bound and the energy distance $|\lambda+E|=0.345~$MeV from the threshold is not very small.

From the above analysis, we can see that the large rms radius for $^{39}$Mg leading to the odd-even staggering is mainly contributed from the blocked weakly bound quasiparticle state $2p_{3/2}$, where the quasiparticle energy $E$ is very close to the threshold energy $-\lambda$ for continuum.

\subsection{Blocking effect and competition between $1f_{7/2}$ and $2p_{3/2}$ }
\label{ResultB}

If we neglect the pair correlation, the last neutron should occupy $1f_{7/2}$ orbit in $^{36-40}$Mg, and $2p_{3/2}$ orbit in $^{41-42}$Mg. Our calculations for $^{37}$Mg with SLy4 parameter show that the last odd neutron occupies the quasiparticle state $1f_{7/2}$, which is in accordance with the configuration without pairing. In the calculation for $^{39}$Mg, on the other hand, the configuration is the one blocking the $2p_{3/2}$ state. This differs from a naive expectation in the single-particle picture since the last odd neutron of $^{39}$Mg would occupy $1f_{7/2}$ orbit if the pairing were neglected. Next, we will explain why the ground state of $^{39}$Mg corresponds to blocking the $2p_{3/2}$ quasiparticle state, and for this we compare the two cases with blocking the $2p_{3/2}$ and $1f_{7/2}$ states.

\begin{table}[!ht]
\tabcolsep=5pt
\caption{Total energy $E_{\rm tot.}$, neutron continuum threshold energy $-\lambda$, quasiparticle energies $E$ for states $2p_{3/2}$ and $1f_{7/2}$, neutron pairing energy $E_{\rm pair}^{\rm n}$, and single-particle energies $\varepsilon$ of orbits $2p_{3/2}$ and $1f_{7/2}$ for the configurations with blocking $1f_{7/2}$ and $2p_{3/2}$ states respectively. Here the box-discretized calculation is employed with the Skyrme parameter SLy4. The unit of the energy is MeV.}
\label{Tab2}
\begin{tabular}{ccc}  \hline\hline
block state &$1f_{7/2}$ &$2p_{3/2}$ \\ \hline
$E_{\rm tot.}$         &-262.761 &-263.075\\
$-\lambda$             &   1.088 &   1.551\\
$E(2p_{3/2})$          &   1.013 &   1.476\\
$E(1f_{7/2})$          &   1.599 &   1.623\\
$E_{\rm pair}^{\rm n}$ &  -4.020 &  -7.211\\
$\varepsilon(2p_{3/2})$&  -0.364 &  -0.362\\
$\varepsilon(1f_{7/2})$&  -2.308 &  -2.211\\ \hline\hline
\end{tabular}
\end{table}

Table~\ref{Tab2} lists the total energy $E_{\rm tot.}$, the neutron continuum threshold energy $-\lambda$, the quasiparticle energies $E$, the neutron pairing energy $E_{\rm pair}^{\rm n}$, and the single-particle energies $\varepsilon$ of $^{39}$Mg calculated with the box-discretized Skyrme-HFB approach with blocking $1f_{7/2}$ and $2p_{3/2}$ states respectively.

From Table~\ref{Tab2}, we can see that $^{39}$Mg is more bound with blocking the $2p_{3/2}$ state compared with blocking the $1f_{7/2}$ state. We notice that if the $2p_{3/2}$ state is blocked, the obtained neutron pairing energy is $E_{\rm pair}^{\rm n}=-7.211~$MeV while $E_{\rm pair}^{\rm n}=-4.020~$MeV with blocking the $1f_{7/2}$ state. The difference in the neutron pairing energies (about $3$~MeV) by blocking the $1f_{7/2}$ and $2p_{3/2}$ states compensates the gap ($\sim$ $2$~MeV) between the $2p_{3/2}$ and $1f_{7/2}$ single-particle orbits. As a consequence, the total energies of these two blocking configurations becomes comparable. Furthermore, $^{39}$Mg is unbound if the last neutron occupies the $1f_{7/2}$ state because the one-quasiparticle state $1f_{7/2}$ is in the continuum with $E>-\lambda$.

\begin{figure}[!ht]
 \includegraphics[width=0.45\textwidth]{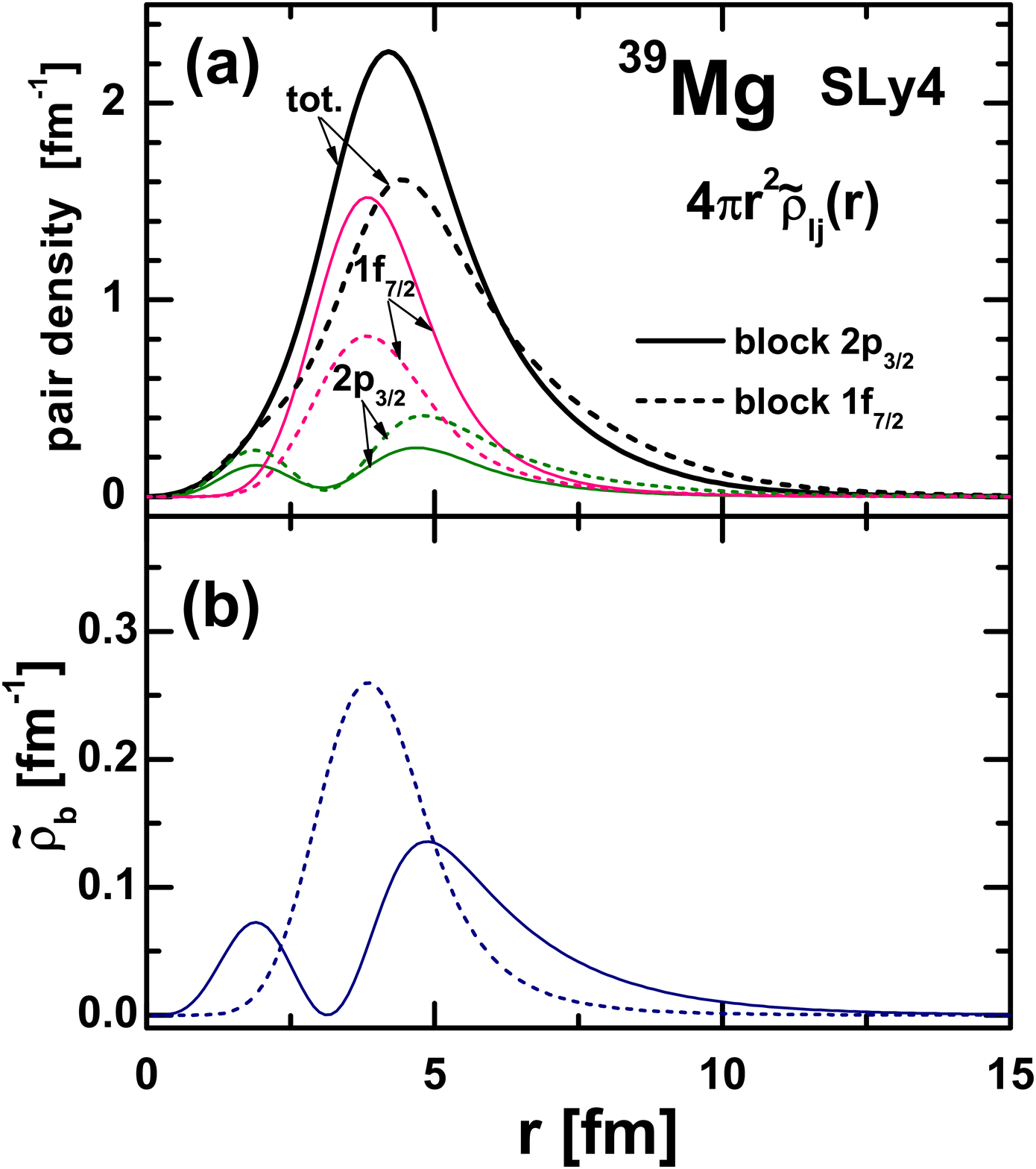}
 \caption{(Color online) (a) Total neutron pair density $4\pi r^{2}\tilde{\rho}(r)$ and neutron pair densities $4\pi r^{2}\tilde{\rho}_{lj}(r)$ for $2p_{3/2}$ and $1f_{7/2}$ states, and (b) the blocking term $\tilde{\rho}_{\rm b}(r)=2 \varphi_{1,n_{\rm b}l_{\rm b}j_{\rm b}}(r)\varphi_{2,n_{\rm b}l_{\rm b}j_{\rm b}}(r)$ for neutron pair density in Eq.~(\ref{EQ:rhot}) for $^{39}$Mg obtained with the box-discretized Skyrme-HFB calculation with blocking quasiparticle states $2p_{3/2}$ (the solid curves) and $1f_{7/2}$ (the dashed curves) respectively.}
 \label{Fig5}
\end{figure}

To analyze the effect of the blocking on the neutron pairing energy $E_{\rm pair}^{\rm n}$, we show in Fig.~\ref{Fig5}(a) the total neutron pair density $4\pi r^{2}\tilde{\rho}(r)$ and the neutron pair densities $4\pi r^{2}\tilde{\rho}_{lj}(r)$ for the blocked $2p_{3/2}$ and $1f_{7/2}$ states, obtained in the box-discretized Skyrme-HFB calculation. The solid curves denote the results with blocking the quasiparticle state $2p_{3/2}$ while the dashed curves are the densities obtained with blocking the $1f_{7/2}$ state. We find that the neutron pair density for the quasiparticle state $1f_{7/2}$ is significantly reduced if it is blocked. The same mechanism is applied to the configuration with blocking $2p_{3/2}$. However, the total neutron pair density with blocking $2p_{3/2}$ is much bigger than that obtained with blocking $1f_{7/2}$. This can be explained in terms of the difference in relative positions of the single-particle orbits and the Fermi surface. From Table~\ref{Tab2} and Fig.~\ref{Fig3}, we can see that the single-particle orbit $2p_{3/2}$ lies farther from the Fermi surface compared with the orbit $1f_{7/2}$. So the blocking term $\tilde{\rho}_{\rm b}(r)=2\varphi_{1,n_{\rm b}j_{\rm b}l_{\rm b}}(r)\varphi_{2,n_{\rm b}j_{\rm b}l_{\rm b}}(r)$ of the pair density of the $2p_{3/2}$ state is smaller than that of $1f_{7/2}$ state, as is shown in Fig.~\ref{Fig5}(b).
With smaller reduction by the blocking term to the pair density, the configuration with blocking the $2p_{3/2}$ state
provides larger pairing energy compared with that blocking the $1f_{7/2}$ state, and
thus makes the nucleus more bound.

From the above analysis, we find that larger neutron pairing energy $|E_{\rm pair}^{\rm n}|$ is obtained for the configuration with blocking the $2p_{3/2}$ quasiparticle state compared with the configuration with blocking $1f_{7/2}$. Furthermore, the difference of neutron pairing energies with blocking the $1f_{7/2}$ and $2p_{3/2}$ states compensates the gap between $2p_{3/2}$ and $1f_{7/2}$ single-particle orbits, leading to the smaller total energy for $^{39}$Mg with blocking $2p_{3/2}$.

\subsection{Odd-even staggering with different Skyrme functionals}
\label{ResultC}

Because of the competition between the two blocking configurations, we can expect that the results may be sensitive to details in the single-particle spectra, especially the gap between orbits $1f_{7/2}$ and $2p_{3/2}$. Let us examine this sensitivity by performing calculations with different Skyrme functionals, which can provide different single-particle spectra.

\begin{figure}[!ht]
 \includegraphics[width=0.45\textwidth]{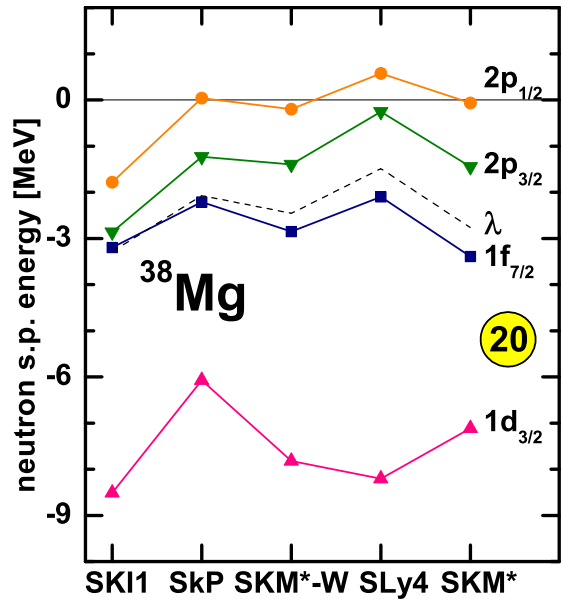}
 \caption{(Color online) Neutron Hartree-Fock (HF) single-particle energy $\varepsilon$ around the Fermi energy for $^{38}$Mg obtained with different Skyrme parameters SKI1, SKP, SKM$^*$-W, SLy4, and SKM$^*$. The dashed line denotes the Fermi energy $\lambda$. The parameter SKM$^*$-W is the same as SKM$^{*}$ except that we decrease the spin-orbit parameter $W_{0}$ by $15\%$.}
 \label{Fig6}
\end{figure}

\begin{figure}[!ht]
 \includegraphics[width=0.45\textwidth]{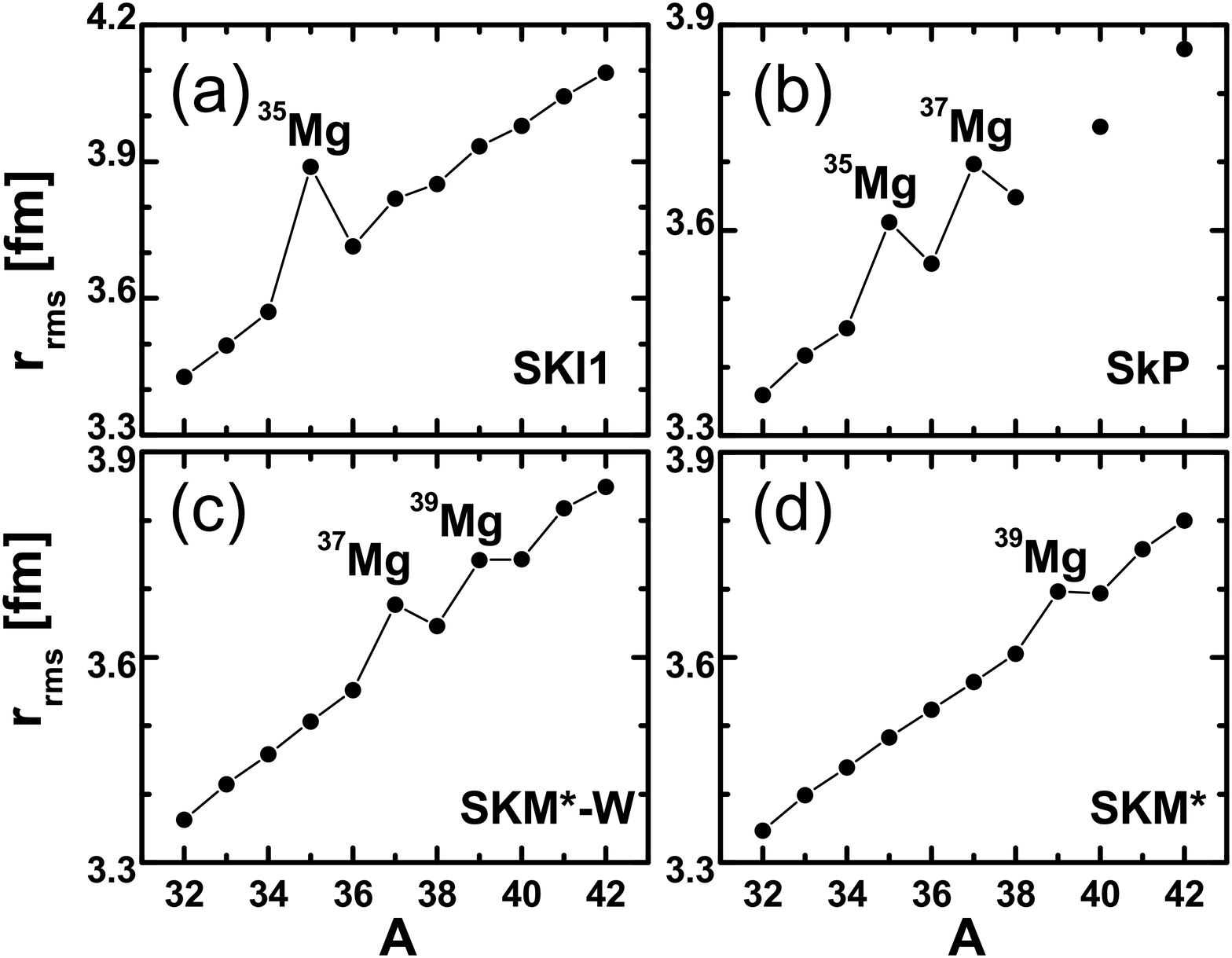}
 \caption{Neutron rms radii for Mg isotopes by the continuum Skyrme-HFB theory with different Skyrme parameters SKI1, SKP, SKM$^*$-W, and SKM$^*$. Note that $^{39}$Mg and $^{41}$Mg are unbound for SkP.}
 \label{Fig7}
\end{figure}

In Fig.~\ref{Fig6}, we show the neutron Hartree-Fock (HF) single-particle energy $\varepsilon$ around the Fermi energy for $^{38}$Mg with different Skyrme parameters SKI1~\cite{NPA1995Reinhard584}, SkP~\cite{NPA1984Doba422}, SKM$^*$-W, SLy4~\cite{NPA1998ChabanatM635Skyrme}, and SKM$^{*}$~\cite{NPA1982BarrelJ386Skms}. The parameter SKM$^*$-W is the same as SKM$^{*}$ except that we decrease the spin-orbit parameter by $15\%$ rather arbitrarily ($W_{0}$ changes from 130 MeV$\cdot$fm$^5$ to 110 MeV$\cdot$fm$^5$). The dashed line denotes the Fermi energy $\lambda$. With these parameters, different single-particle spectra are obtained. Especially, the gap between orbits $2p_{3/2}$ and $1f_{7/2}$ increases from SKI1 to SKM$^*$. As a result, the energy competition between the two configurations with blocking $2p_{3/2}$ and $1f_{7/2}$ varies for different parameters.

Figure~\ref{Fig7} shows different patterns of odd-even staggering obtained in the calculations with the parameters SKI1, SkP, SKM$^*$-W, and SKM$^*$. We find that different Skyrme functionals provide the ground state with different blocking configurations. For SKI1, the gap between $2p_{3/2}$ and $1f_{7/2}$ orbits is the smallest compared with other parameters and it is easier to obtain a smaller total energy with blocking the $2p_{3/2}$ state. The odd-even staggering appears at $^{35}$Mg as shown in Fig.~\ref{Fig7}(a) since the configuration $2p_{3/2}$ becomes the ground state. From the parameter SkP to SKM$^*$, the gap between $2p_{3/2}$ and $1f_{7/2}$ orbits increases, and the ground state configuration for odd$-A$ nuclei changes from blocking the state $2p_{3/2}$ to $1f_{7/2}$. The first odd Mg isotope with blocking the $2p_{3/2}$ state is $^{35}$Mg for SkP, $^{37}$Mg for SKM$^{*}-$W,  $^{39}$Mg for SLy4, and $^{39}$Mg for SKM$^{*}$.

It is noted that the strong odd-even staggering only appears at $^{35}$Mg with SKI1 parameter. This is because that the obtained single-particle orbit $2p_{3/2}$ has a relatively large binding energy $\epsilon\sim -3~$MeV. As a result, although the one-quasiparticle state of $^{37}$Mg and $^{39}$Mg is $2p_{3/2}$, the radii are not very large and only very weak odd-even staggering appears. This mechanism is also applied to $^{41}$Mg with SKM$^{*}$-W in panel (c), and $^{41}$Mg with SKM$^{*}$ in panel (d).

From the above analysis, we demonstrate that the ground state configurations of the neutron-rich odd$-A$ Mg nuclei are sensitive to the details in the single-particle spectra, especially the gap between orbits $1f_{7/2}$ and $2p_{3/2}$. If the gap between $1f_{7/2}$ and $2p_{3/2}$ is small, it is easier to make the nucleus more bound with blocking $2p_{3/2}$ orbit due to the pairing correlation.
As a result, different patterns of odd-even staggering appear at different nuclei obtained with different
Skyrme functionals.

\section{CONCLUSIONS}\label{Chapter5}

The self-consistent continuum Skyrme-HFB theory for odd-$A$ nuclei formulated with the Green's function technique in the coordinate space is developed by incorporating the blocking effect. With the present theory, we predict odd-even staggering in the neutron rms radii in neutron-rich Mg isotopes.

The calculation performed with the SLy4 parameter shows that odd-even staggering of the neutron rms radius appears in $^{39}$Mg with the ground state configuration occupying the $2p_{3/2}$ state. The large rms radius is mainly contributed from the blocked weakly bound quasiparticle state $2p_{3/2}$, which is located very close to the threshold for the unbound continuum states. The total energy of the $2p_{3/2}$ configuration is lower than that of the $1f_{7/2}$ configuration although the latter is expected to be lower if the pairing is neglected. This is because the difference of the pairing energies caused by the blocking effect overcomes the gap between $2p_{3/2}$ and $1f_{7/2}$ single-particle orbits.

Furthermore, we also studied the odd-even staggering with different Skyrme parameters and we find that the ground state configurations for the odd$-A$ Mg isotopes with $A\geq 35$ are sensitive to the details in the single-particle spectra, especially the gap between orbits $1f_{7/2}$ and $2p_{3/2}$. If the gap between $1f_{7/2}$ and $2p_{3/2}$ is small, it's more easier to obtain the ground state with blocking $2p_{3/2}$ state for the Mg isotopes. Also, odd-even staggering appears for different nuclei with different parameters.

\begin{acknowledgments}
The authors thank T. Ohtsubo for the fruitful discussions on the experimental data, and K. Yoshida and L. L. Li for very helpful advises and suggestions on theory. This work was partly supported by the Major State 973 Program of China No.~2013CB834400, Natural Science Foundation of China under Grants No.~11175002 and No.~11335002, the Research Fund for the Doctoral Program of Higher Education under Grant No.~20110001110087, and Grants-in-Aid for Scientific Research (No.~21340073, No.~23540294 and No.~24105008) from the Japan Society for the Promotion of Science.
\end{acknowledgments}


\end{document}